\documentclass[twocolumn,showpacs,preprintnumbers,
amsmath,amssymb,floatfix,nofootinbib]{revtex4-2}

\usepackage{dcolumn}
\usepackage{bm}
\usepackage{amsthm}
\usepackage{amsfonts}

\usepackage{anysize}

\usepackage{amsmath}
\usepackage{mathtools}
\usepackage[english]{babel}
\usepackage{graphicx}

\usepackage{caption}
\usepackage{subcaption}

\usepackage{amssymb}

\newcommand{\be}{\begin{equation}}
\newcommand{\ee}{\end{equation}}
\newcommand{\ba}{\begin{eqnarray}}
\newcommand{\ea}{\end{eqnarray}}
\newcommand{\non}{\nonumber}

\newcommand{\ce}{{\cal{E}}}
\newcommand{\cl}{{\cal{L}}}
\newcommand{\ck}{{\cal{K}}}

\newcommand{\nin}{\noindent}

\usepackage{color}

\begin{document}

\title{Tilted Circular Orbits around a Kerr Black Hole}
\author{A. M. Al Zahrani}
\email{ama3@ualberta.ca
\newline amz@kfupm.edu.sa}
\affiliation{Physics Department, KFUPM, Dhahran 31261, SA}

\begin{abstract}
  We study circular orbits that are tilted with respect to the equatorial plane around a Kerr black hole. We write the equations for the parameters of a tilted circular orbit in terms of the orbit's radius and the Carter constant, or equivalently, the tilt angle. The tilted innermost stable circular orbits (TISCO)s are discussed as well as the last circular orbits. The azimuthal precession of an orbit is then studied and an approximate expression for the precession speed is given. We finally try to link tilted circular orbits to quai-periodic oscillations in some astrophysical black hole systems.
\end{abstract}


\maketitle

\section{Introduction}
Astrophysical black holes can have bright accretions accretion disks and jets. The plane of an accretion disk is usually considered to be perpendicular to the black hole's spin, although accretion disks are very likely to be tilted. According to Ref.~\cite{FMW}, it is likely that, if not most, many X-ray binaries form with misaligned angular momenta. More importantly, there are abundant observational evidences for the existence of tilted accretion disk black hole systems. For example, the exotic system SS 433 contains a black hole with a percessing accretion disk and jet~\cite{Fa}. The galaxy NGC 4258 (M106) has an AGN with a warped disk~\cite{HGM}. 

Tilted accretion disk models have been proposed to explain the emission variability from some black hole systems. In particular, they were used to explain the quasi-periodic oscillations in X-ray binaries~\cite{AM}. The effect of the Lense-Thirring precession on a tilted accretion disk around a Kerr black hole was studied in Ref.~\cite{BP}. It was concluded that the inner part of the disk will get aligned with the black hole's spin. This effect is known as the Bardeen–Petterson effect. The derivation of the basic equations of twisted accretion disks with applications to X-ray binaries was given in Refs.~\cite{P1,P2}. The equations governing the time-dependent structure of a twisted thin accretion disk and their properties were later given in Ref.~\cite{P3}. The equations in Refs.~\cite{P1,P2,P3} were revised and corrected in Ref.~\cite{HBS}. Later on, it was demonstrated that the twist evolution equations derived previously were incorrect as the angular momentum was not conserved in them~\cite{PP}. About a decade later, a simple set of equations that governs the time evolution of a twisted accretion disk was given in Ref.~\cite{Pr}. These equations were then modified by adding an effective term corresponding to the Lense-Thirring precession~\cite{SF}, and then solved analytically for a warped accretion disk around a slowly spinning black hole~\cite{CB}.    

The physics of tilted accretion disks is still not well understood. Several simulations of tilted accretion disks have been performed to demystify the physics behind them. The first fully general relativistic three-dimensional
hydrodynamic numerical studies of tilted thick-disk accretion onto rapidly rotating black holes were performed in Ref.~\cite{FA}. It was found that the Lense-Thirring precession caused the disk to warp but only within a specific radius in the disk. A consecutive numerical simulation which
fully considers the effects of the black hole spacetime curvature along with the magnetorotational turbulence was done in Ref.~\cite{Fr}. It was found that accretion onto the black hole occurs mainly through two opposing plunging streams that start from high latitudes with respect to both the black hole and disk midplanes. More importantly, it was found that the main body of the disk remains tilted with respect to the symmetry plane of the black hole, unlike what would be expected from the Bardeen-Petterson effect. The spcetime precession causes a global precession of the main disk body that has a frequency of $3(M_\odot/M)$ Hz. In ref.~\cite{PM}, simulations showed that the disk spin and black hole spin alignment can occur by the Blandford-Znajec jet torque before the Lense-Thirring torque becomes important. This is because disc material get aligned by the Blandford-Znajec jet torque before the Lense-Thirring torque that falls steeply with radius takes action. High resolution 3D general relativistic magnetohydrodynamic simulations of tilted thick accretion discs around rapidly spinning black holes was performed~\cite{L1}. It was concluded that these accretion disks generate relativistic jets that propagate along the disk axis, not the black hole spin axis. A similar numerical simulation for thin tilted accretion disks was performed in Ref.~\cite{L2}. It showed that the inner part of the disk undergoes Bardeen-Peterson alignment. Furthermore, other simulations revealed that the disk lauches powerful relativistic jets along the angular momentum vector of the outer tilted part of the accretion disk. The simulations in Ref.~\cite{WQB} found that tilted disks quickly reach a warped and twisted shape that rigidly precesses about the
black hole spin axis and magnetized polar outflows form along the disk rotation axis.

In this paper we try to enhance the understanding of the physics of tilted accretion disks by studying their most primitive building blocks; tilted circular orbits. We start by reviewing the equations of motion of particles in Kerr spacetime and then study tilted circular orbits in Sec.~\ref{s2}. In Sec.~\ref{s3}, we solve the equations of motion for a tilted circular orbit numerically and study the precession of tilted circular orbits. We give an approximate expression for the precession angular speed in Sec.~\ref{s4}. Finally, a summary of the main findings is given in Sec.~\ref{sum}. 

We use the sign conventions adopted in Ref.~\cite{MTW} and geometrical units where $c=G=1$.

\section{Tilted Circular Orbits} \label{s2}

\subsection{Equations of motion}

We start by reviewing the dynamics of massive particles in the Kerr spacetime. The spacetime geometry around a rotating black hole is described by the Kerr metric. For a black hole of mass $M$ and spin angular momentum $J=aM$, the Kerr metric in Boyer-Linquist coordinates reads~\cite{FN}
\begin{eqnarray}
ds^2=&-&\Sigma\frac{\Delta}{A}dt^2+\frac{\Sigma}{\Delta}dr^2+\Sigma d\theta^2\nonumber \\
&+&\frac{A}{\Sigma}\left(d\phi-\frac{2aMr}{A}dt\right)^2\sin^2\theta,
\end{eqnarray}
where
\begin{eqnarray}
\Sigma &=&r^2+a^2\cos^2\theta, \:\:\: \Delta=r^2+a^2-2Mr, \non \\
&&\:\:\:A=(r^2+a^2)^2-a^2\Delta \sin^2\theta,
\end{eqnarray}

\nin and $a$, with $-M\leq a\leq M$, is the rotation parameter.

The Kerr spacetime admits two commuting Killing vectors
\begin{equation}
\xi^{\mu}_{(t)}=\delta^{\mu}_t, \:\:\: \xi^{\mu}_{(\phi)}=\delta^{\mu}_{\phi},
\end{equation}

\noindent and a Killing tensor
\begin{equation}
K^{\mu\nu}=\Delta k^{(\mu}l^{\nu)}+r^2g^{\mu\nu},
\end{equation}

\noindent where
\begin{eqnarray}
l^{\mu}&=&\frac{1}{\Delta}\left[(r^2+a^2)\delta_t^{\mu}+\Delta \delta_r^{\mu}+a \delta_{\phi}^{\mu}\right], \\
k^{\mu}&=&\frac{1}{\Delta}\left[(r^2+a^2)\delta_t^{\mu}-\Delta \delta_r^{\mu}+a \delta_{\phi}^{\mu}\right].
\end{eqnarray}

\noindent as described in Ref.~\cite{Poi}. Consider a particle in the Kerr spacetime moving with four-velocity $u^{\mu}$. The three Killing symmetries are associated with three constants of the particle's motion
\begin{eqnarray}
-\ce &=& p_{\mu}\xi^{\mu}_{(t)}/m,\\
\cl &=& p_{\mu}\xi^{\mu}_{(\phi)}/m,\\
\ck &=& u_{\mu}u_{\nu}K^{\mu\nu},
\end{eqnarray}

\noindent where $p^{\mu}=mu^{\mu}$ is the particle's four-momentum. $\ce$ and $\cl$ are the specific energy and azimuthal angular momentum, respectively, and $\ck$ is the Carter constant introduced in Ref.~\cite{Car}. Using these three constants of motion along with the normalization $u_{\mu}u^{\mu}=-1$ we reduce the equations of motion to quadratures:
\begin{eqnarray}
\dot{t}&=&\ce+\frac{2Mr[(r^2+a^2)\ce-a\cl]}{\Delta\Sigma},\label{tdot}\\
\dot{\phi}&=&\frac{\cl}{\Sigma\sin^2\theta}+\frac{a(2Mr\ce-a\cl)}{\Delta\Sigma}\label{phidot},\\
\Sigma^2\dot{r}^2&=&[(r^2+a^2)\ce-a\cl]^2-\Delta(r^2+\ck),\label{rdot}\\
\Sigma^2\dot{\theta}^2&=&\ck-a^2\cos^2{\theta}-\left(a\ce\sin{\theta}-\frac{\cl}{\sin{\theta}}\right)^2\label{thethadot},\hspace{5mm}
\end{eqnarray}

\noindent where the overdot denotes the derivative with respect to the proper time of the particle. The dynamics is invariant under reflection with respect to the equatorial plane
\begin{equation}\label{sym1}
\theta\rightarrow\pi-\theta, \hspace{1cm} \dot{\theta}\rightarrow-\dot{\theta}.
\end{equation}
It is also invariant under the transformations
\begin{equation}\label{sym2}
\phi\rightarrow-\phi, \hspace{3mm} \dot{\phi}\rightarrow-\dot{\phi}, \hspace{3mm} \cl\rightarrow-\cl, \hspace{3mm} a\rightarrow -a.
\end{equation}
There are two dynamically distinct modes of motion, depending on whether the black hole's spin and particle's azimuthal angular momentum are parallel ($a\cl>0$) or anti-parallel ($a\cl<0$). Without loss of generality, we will keep $\cl$ positive while $a$ can take both signs.

\subsection{Tilted circular orbits}

According to Eq.~(\ref{thethadot}), a tilted orbit will be oscillating about the equatorial plane between $\theta_-=\pi/2-\zeta$ and $\theta_+=\pi/2+\zeta$, where $\zeta$ is the tilt angle with respect to the equatorial plane, if  
\begin{equation}\label{cc2}
\ck = a^2\sin^2\zeta+\left(a\ce\cos\zeta-\frac{\cl}{\cos\zeta}\right)^2.
\end{equation}

Let us define $R(r)$ to be the right hand side of Eq.~(\ref{rdot}):  
\begin{eqnarray}
R(r):=[(r^2+a^2)\ce-a\cl]^2-\Delta(r^2+\ck).\label{rpot}
\end{eqnarray}
$R(r)$ is positive semidefinite; it vanishes at the radial turning points only. Circular orbits exist where $R(r)$ and its first derivative $R'(r)$ vanish. These two conditions yield
\begin{eqnarray}
&&[(r^2+a^2)\ce-a\cl]^2-\Delta(r^2+\ck)=0,\:\:\:\:\:\:\: \\
&&2r\ce[(r^2+a^2)\ce-a\cl]-r\Delta \nonumber\\
&&\hspace{20mm}-(r-M)(r^2+\ck)=0.\:\:
\end{eqnarray}

\nin Solving these equations for $\ce$ and $\cl$ one can obtain explicit expression of the form
\begin{eqnarray}
&&\ce=\ce(r,a,\zeta),\label{ec}\\
&&\cl=\cl(r,a,\zeta). \label{lc}
\end{eqnarray}
The functions $\ce(r,a,\zeta)$ and $\cl(r,a,\zeta)$ are too complicated to be written explicitly. They reduce to the known expressions for equatorial orbits when $\zeta=0$. $\ce$ is positive for all circular orbits of any tilt angle.

\subsection{The tilted innermost stable circular orbits}

\nin A tilted circular orbit is a TISCO when $R''(r)$ vanishes, or
\begin{eqnarray}
&&a\left(a+2\ce\cl-2a\ce^2\right)+\ck \nonumber\\
&&\hspace{5mm}-6r\left[\left(\ce^2-1\right)r+M\right]=0.
\end{eqnarray}

\begin{figure}[h!]
  \centering
  \includegraphics[width=0.5\textwidth]{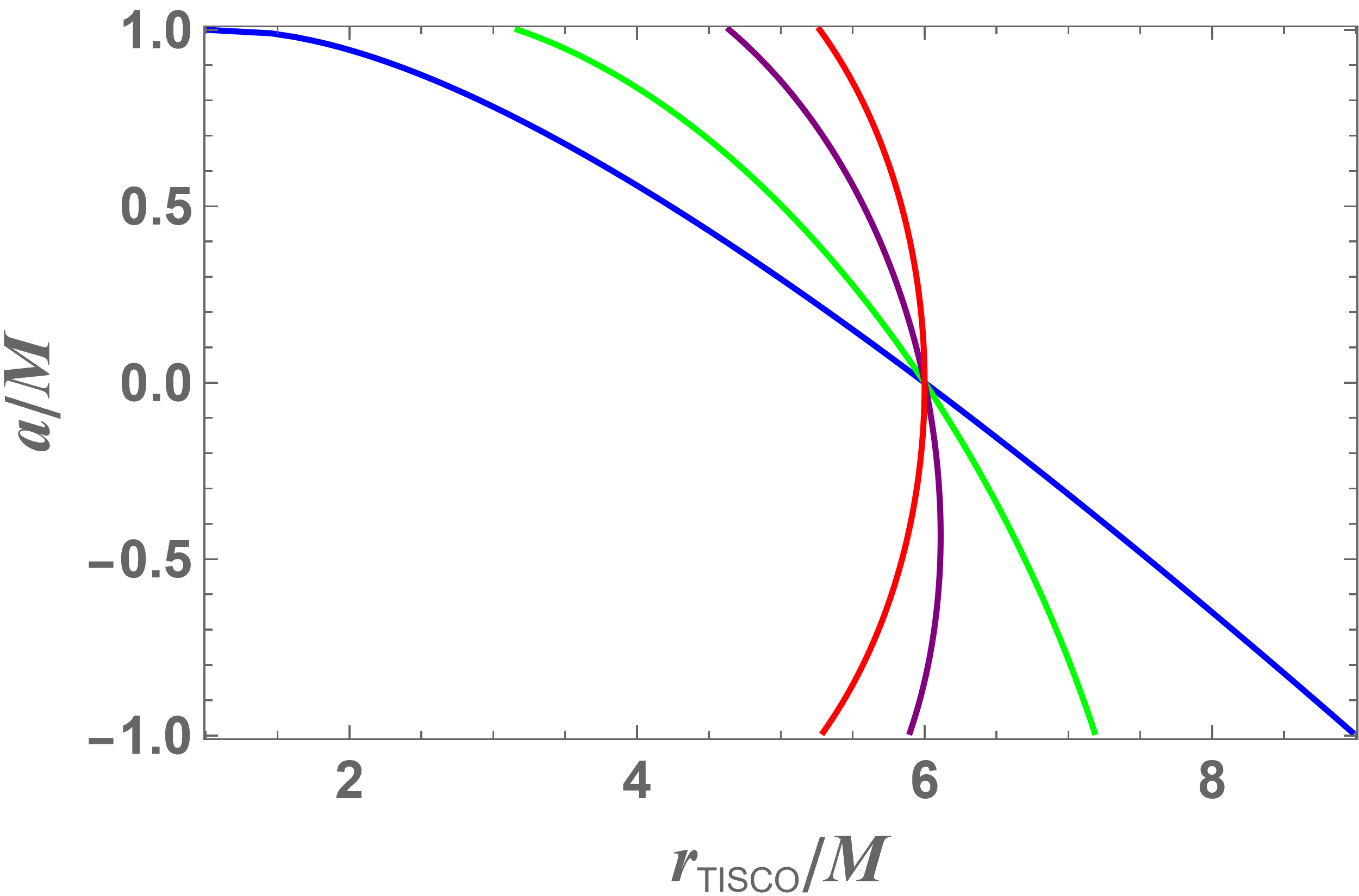}
  \caption{The dependence of the radius of the tilted innermost stable circular orbit $r_{_\text{TISCO}}$ on the black hole's rotation parameter $a$ for three different tilt angles. $\zeta=0$ (blue), $\zeta=\pi/3$ (green), $\zeta=\pi/2$ (red).}
  \label{fig:arms}
\end{figure}

\nin Figure~\ref{fig:arms} shows how the radius of the TISCO  $r_{_{\text{TISCO}}}$ changes with $a$ for different values of $\zeta$. The TISCOs radii lie in the interval $[M,9M]$. In general, as the orbit gets more tilted, $r_{_{\text{TISCO}}}$ gets farther (closer) for positive (negative) $a$. The curves become more symmetric with respect to $a$ as an orbit gets more tilted. When $\zeta=\pi/2$, the curve becomes completely symmetric as it should be. As expected, all curves meet at $r_{_{\text{TISCO}}}=6M$ when $a=0$. There is a critical value of the tilt angle $\zeta_c\approx 71^\circ$ after which the relation between $a$ and $r$ is not one-to-one. A specific value of $r_{_{\text{TISCO}}}$ can correspond to two different values of $a$.

\begin{figure}[h!]
  \centering
  \includegraphics[width=0.5\textwidth]{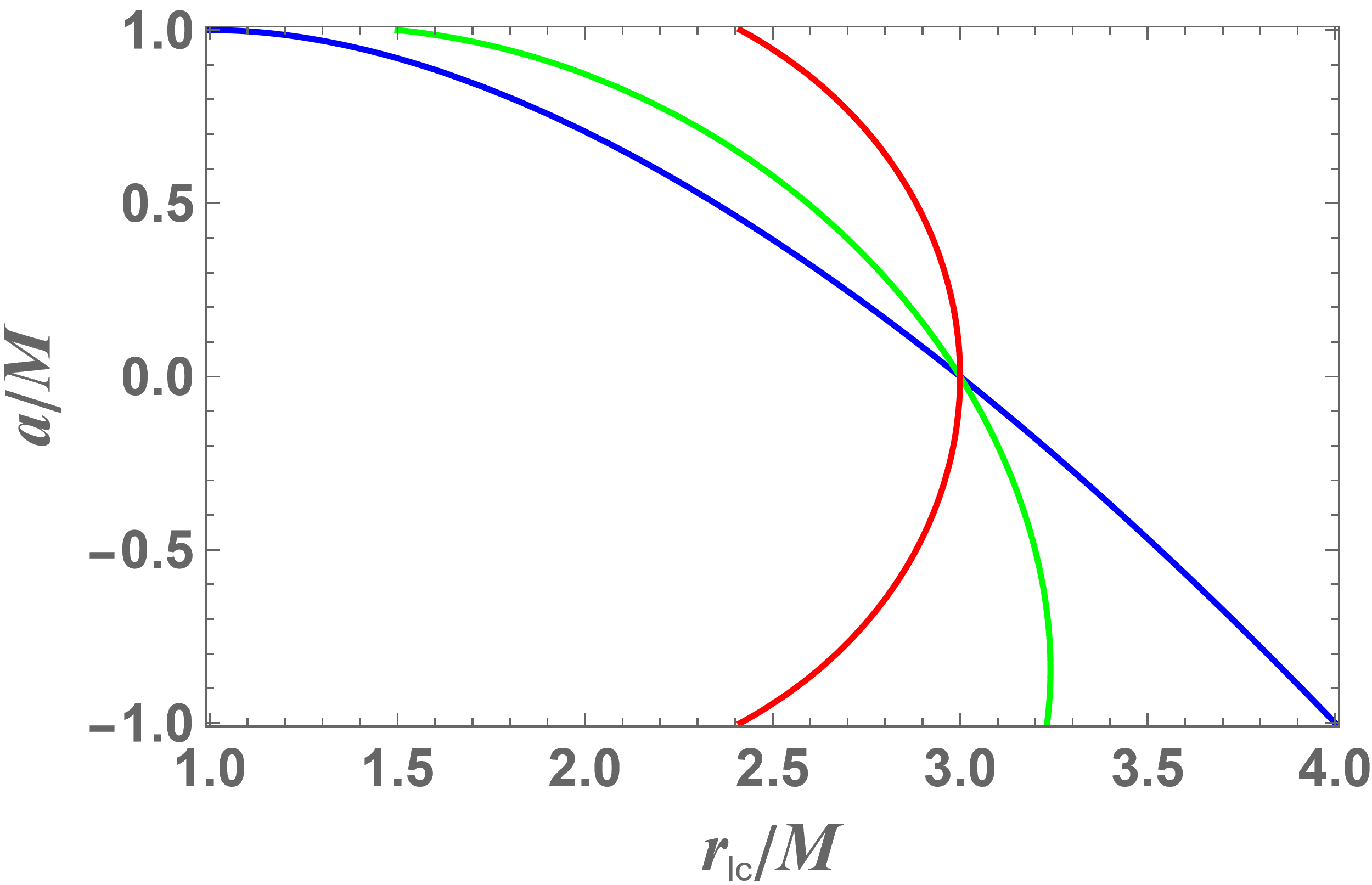}
  \caption{The dependence of the radius of the last circular orbit $r_{_\text{lc}}$ on the black hole's rotation parameter $a$ for three different tilt angles. $\zeta=0$ (blue), $\zeta=\pi/3$ (green), $\zeta=\pi/2$ (red).}
  \label{fig:arlc}
\end{figure}

\nin The radius of the last circular orbit $r_{\text{lc}}$, is obtained by setting $\ce^{-1}=0$. Figure~\ref{fig:arlc} shows how $r_{\text{lc}}$ changes with $a$ for the same three values of $\zeta$ as in Figure~\ref{fig:arms}. The effect of changing $\zeta$ is similar in both figures. Again, all curves meet at $r_{_{\text{lc}}}=3M$ when $a=0$.

\section{The Three-Dimensional Motion of a particle in a tilted circular orbit} \label{s3}

The dynamics of a particle in curved spacetime is governed by the geodesic equation
\begin{equation}\label{de}
m{u}^{\nu}\nabla_{\nu}u^{\mu}=0,
\end{equation}

\begin{figure}[h!]
  \centering
  \includegraphics[width=0.4\textwidth]{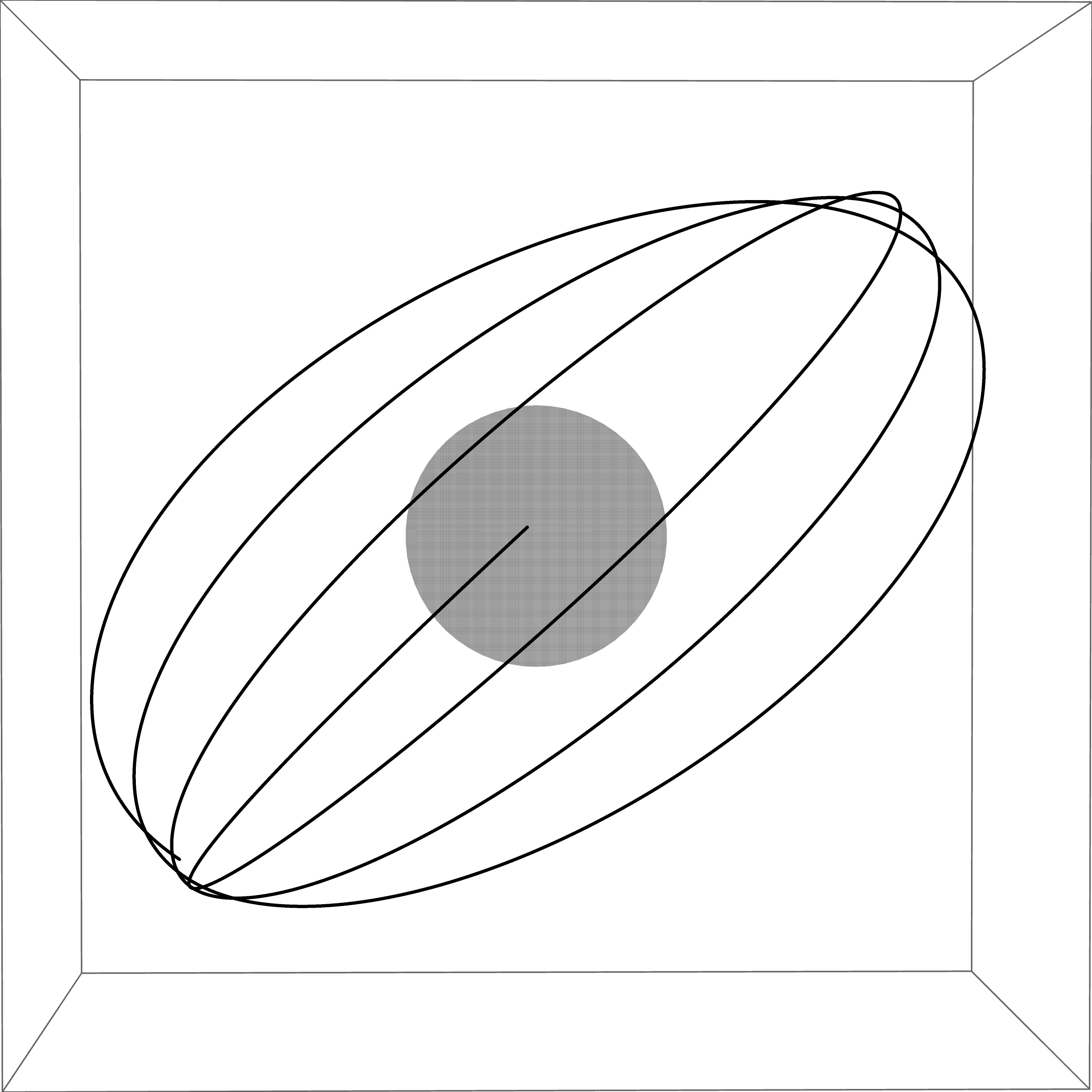}
  \caption{The trajectory of a particle in a circular orbit of radius $r=7$ with a tilt angle of $\zeta=\pi/4$ and black hole's spin $a=M/2$.}\label{fig:tra1}
\end{figure}

\nin where $\nabla_{\mu}$ is the covariant derivative. We can obtain the trajectory of a particle by numerically integrating the $r$ and $\theta$ components of Eq.~(\ref{de}) and then integrating Eq.~(\ref{phidot}). An example is shown in Fig.~\ref{fig:tra1} for a circular orbit with $r=7M$ and tilt angle $\zeta=\pi/4$ with the black hole's spin $a=M/2$ for a few periods.
\begin{figure}[h!] 
  \centering
  \includegraphics[width=0.4\textwidth]{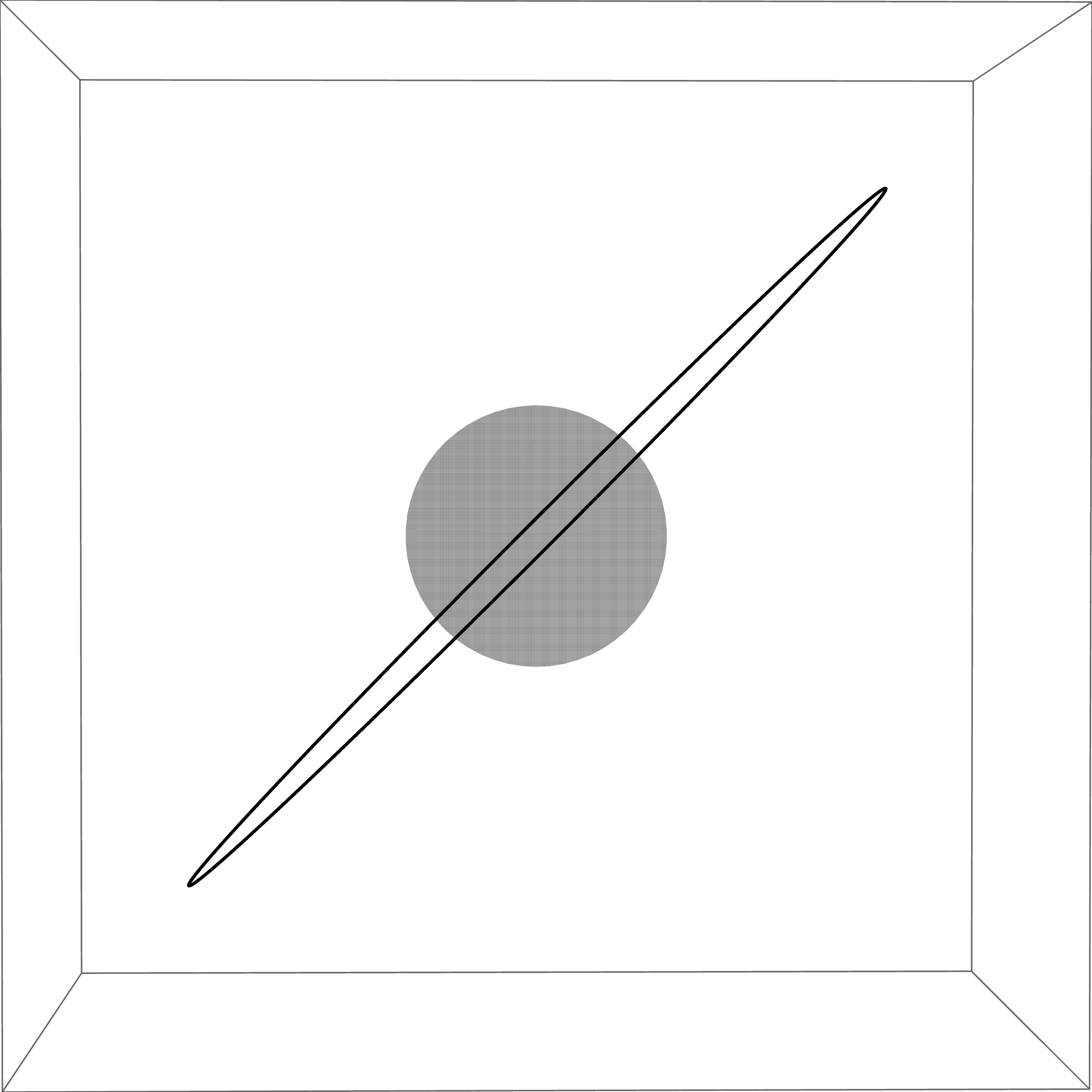}
  \caption{The orbit in Fig.~\ref{fig:tra1} becomes perfectly circular after applying the transformation of Eq.~\ref{phi2} with the right value of $\omega_p$.}\label{fig:tra2}
\end{figure} 
\nin The trajectory looks complicated at first glance. However, a transformation of the form
\begin{equation}
\phi(\tau) \rightarrow \phi(\tau)-\omega_p\tau, \label{phi2}
\end{equation}
makes the orbit still and perfectly circular, as shown in Fig.~\ref{fig:tra2}. Here $\omega_p=\omega_p(r,\zeta,a)$ is the precession angular speed of the circular orbit. Surprisingly, $\omega_p$ is constant over the particle's orbit.

\begin{figure}[h!]
  \centering
  \includegraphics[width=0.5\textwidth]{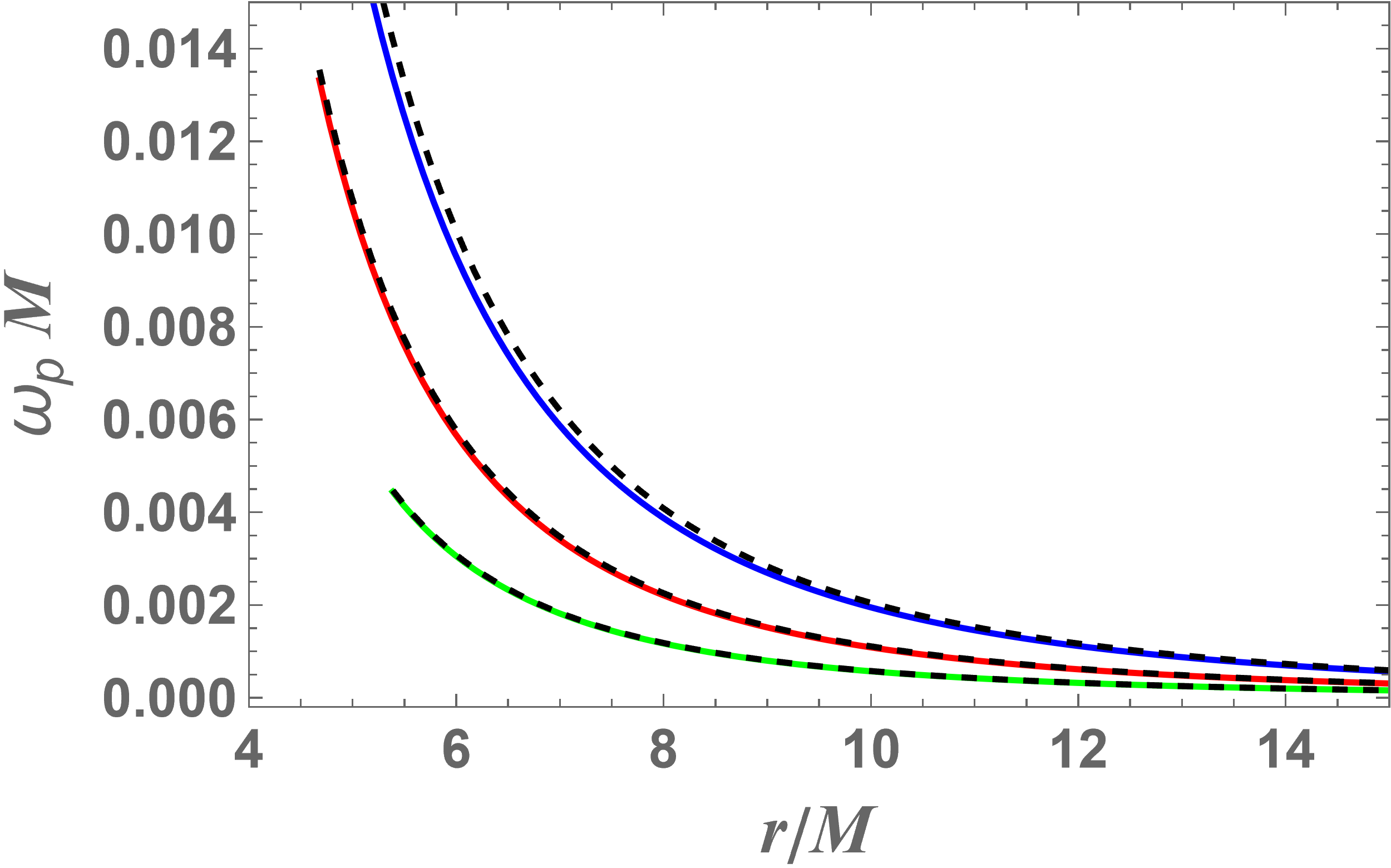}
  \caption{$\omega_p$ vs. $r$ for $a=M$ (blue), $a=M/2$ (red) and $a=M/4$ (green). The blue curve approaches $0.298$ when $r=r_{_{\text{TISCO}}}$. $\zeta=\pi/4$. The dashed curves correspond to the approximate formula (see below).}\label{fig:omegap}
\end{figure}

\begin{figure}[h!]
  \centering
  \includegraphics[width=0.5\textwidth]{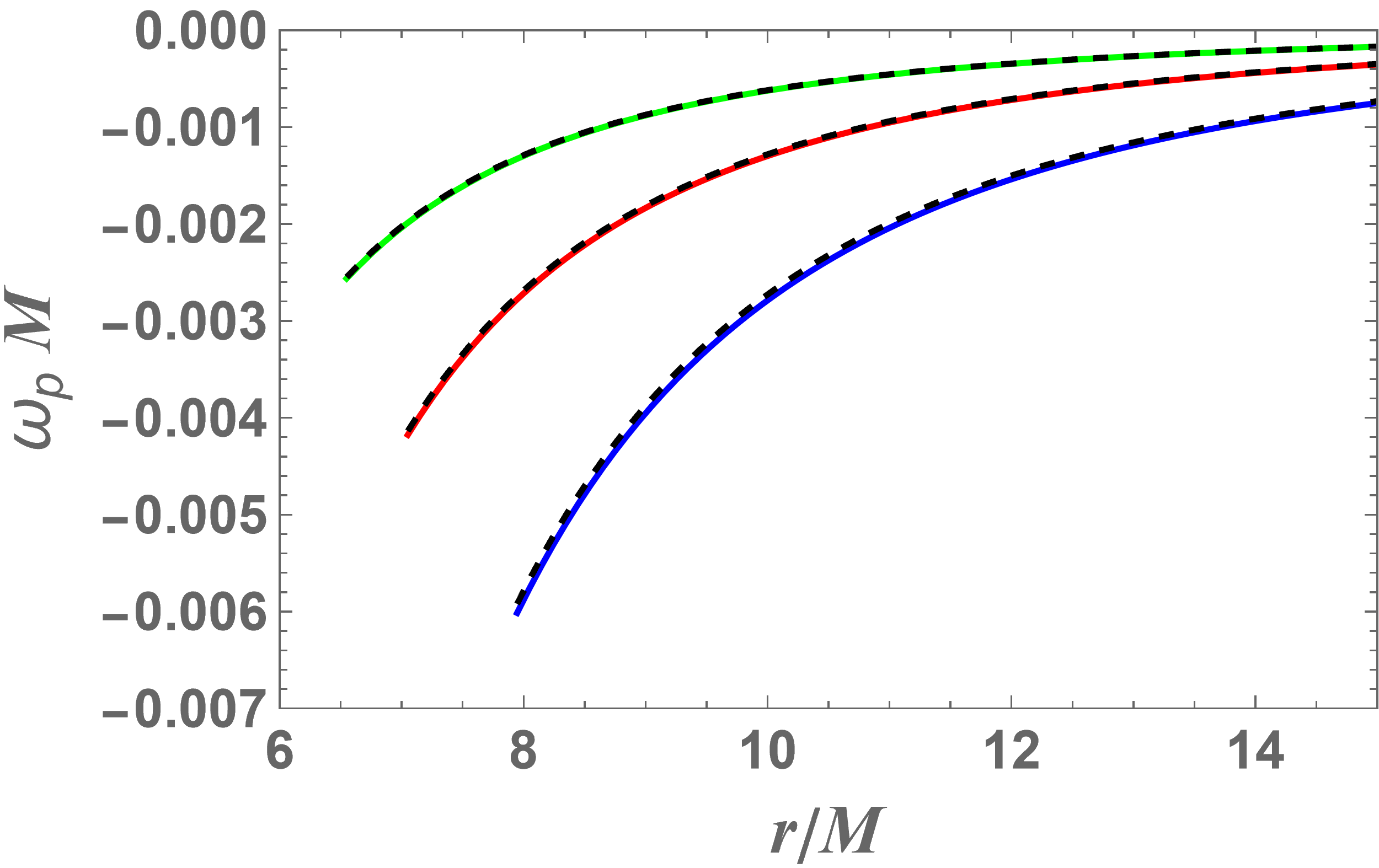}
  \caption{$\omega_p$ vs. $r$ for $a=-M$ (blue), $a=-M/2$ (red) and $a=-M/4$ (green). $\zeta=\pi/4$. The dashed curves correspond to the approximate formula (see below).}\label{fig:omegan}
\end{figure}

It is astrophysically interesting to see how $\omega_p$ varies as $r$, $\zeta$ and $a$ change. Numerical analyses revealed that the effect of changing $\zeta$ on $\omega_p$ is marginal. We will therefore fix $\zeta$ and study the effects of $a$ and $r$ only. Figure~\ref{fig:omegap} shows how $\omega_p$ changes with $r$ for different positive values of $a$. Figure~\ref{fig:omegan} shows how $\omega_p$ changes with $r$ for different negative values of $a$. In both figures, $\zeta=\pi/4$. We can see that $\omega_p$ is directly proportional to $a$. It is positive (negative) when $a$ is positive (negative). Moreover, $\omega_p$ falls quickly as $r$ increases.
\\

\subsection{Precession period}

\begin{figure}[h!]
  \centering
  \includegraphics[width=0.45\textwidth]{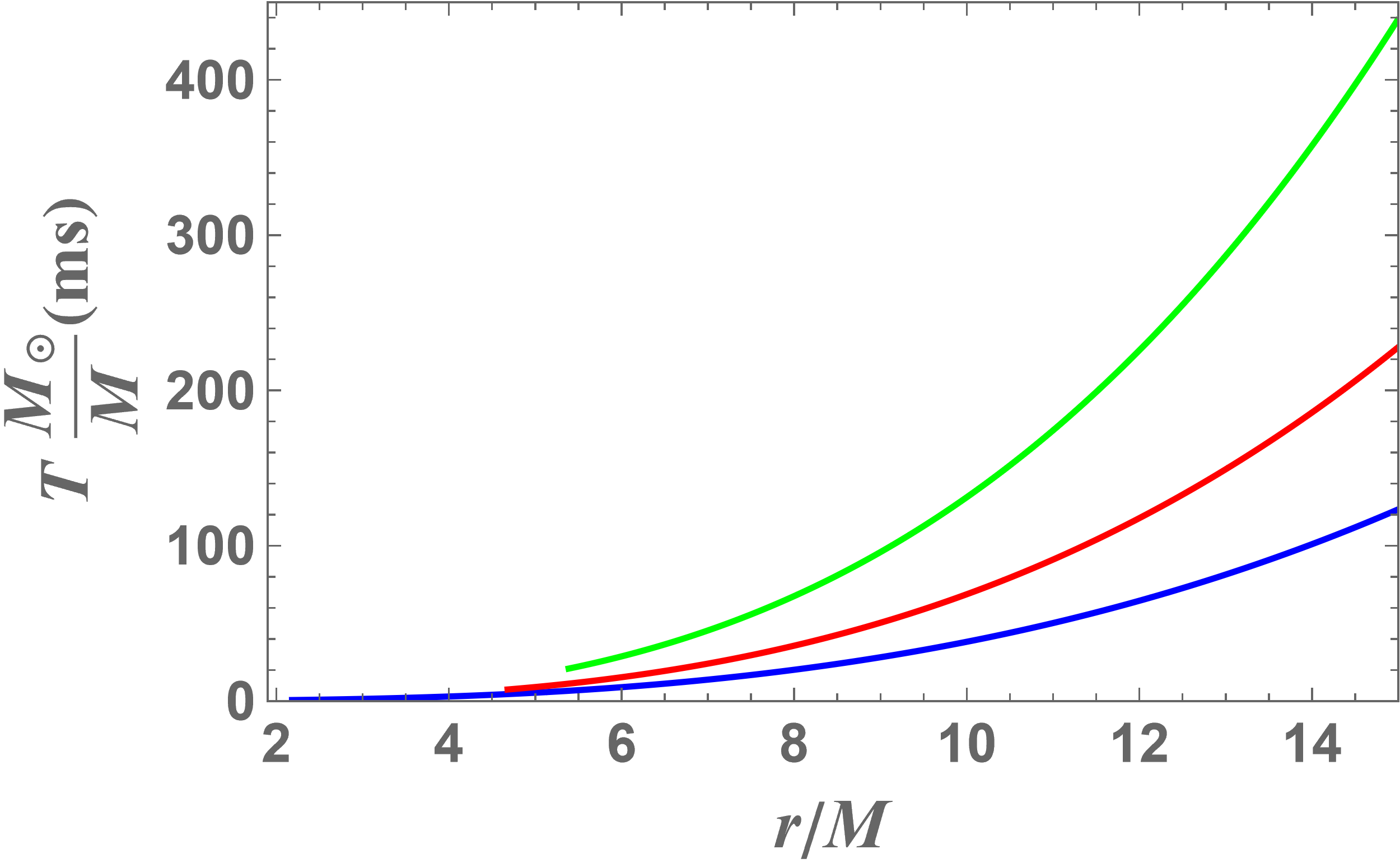}
  \caption{The precession periods as measured by a distant stationary observer corresponding to the values of $\omega_p$ values in Fig.~\ref{fig:omegap}.}\label{fig:periodp}
\end{figure}

\begin{figure}[h!]
  \centering
  \includegraphics[width=0.45\textwidth]{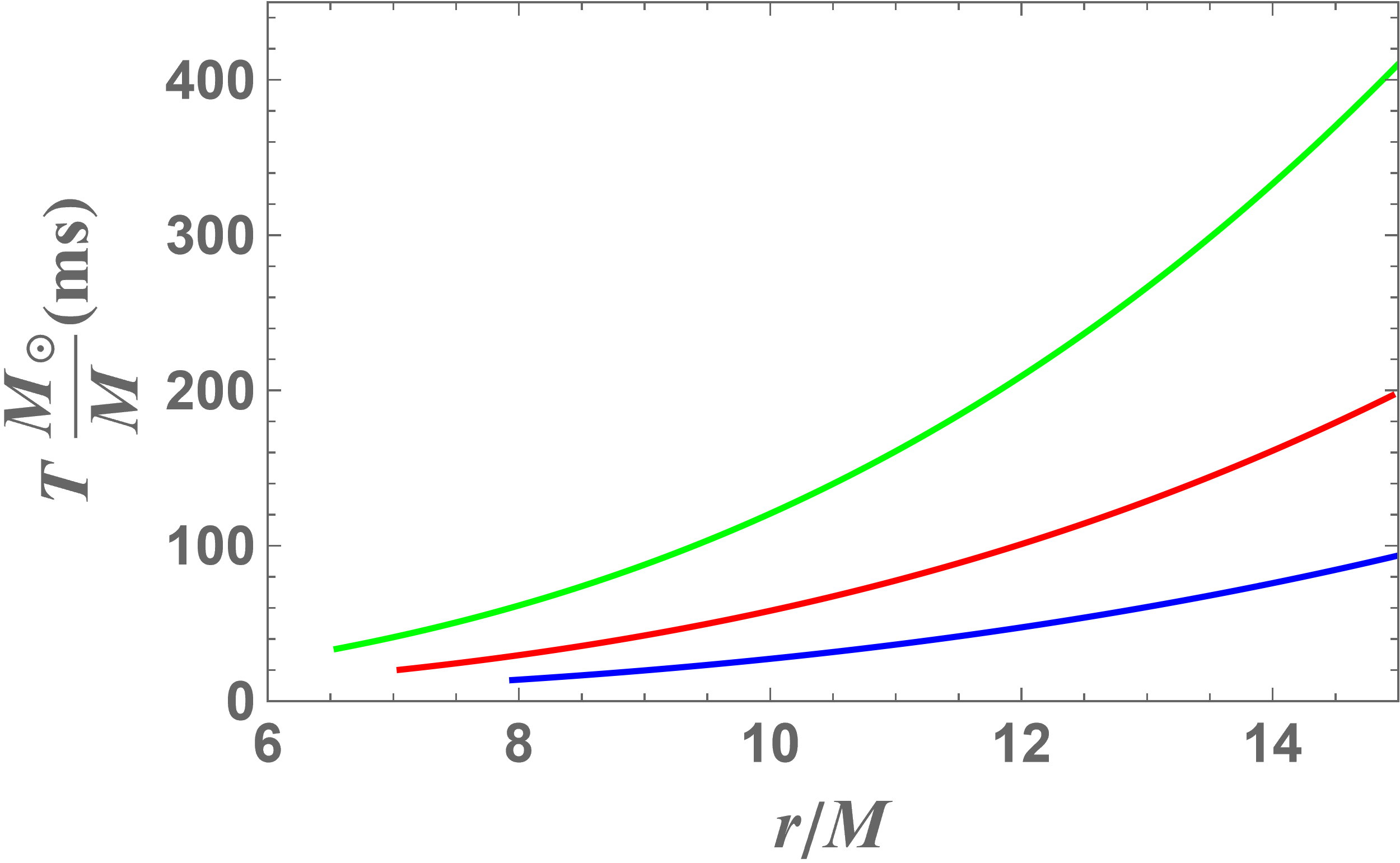}
  \caption{The precession periods as measured by a distant stationary observer corresponding to the values of $\omega_p$ values in Fig.~\ref{fig:omegan}.}\label{fig:periodn}
\end{figure}

As mentioned in the introduction, it was proposed that the variability of black hole systems is due to the precession of their tilted accretion disks. Let us now calculate the precession period of a tilted circular orbit. In conventional units, the precession period (in ms) relative to the orbiting particle $T'$ is given by
\begin{equation}
T' = \frac{2\pi M}{c\omega_p}=\frac{62.8}{\omega_p}\frac{M}{M_\odot}.
\end{equation} 
One can easy calculate the precession period measured by a far away observer at rest $T$ using Eq.~\ref{tdot}. Figures~\ref{fig:periodp}~and~\ref{fig:periodn} show $T$ corresponding to $\omega_p$ values in Figs.~\ref{fig:omegap}~and~\ref{fig:omegan}, respectively. 


The precession period for stellar mass black holes is typically few seconds. For supermassive black holes, the precession period is typically few days to few years. Therefore, both fast and slow variabilities in black hole systems can be attributed to tilted accretion disks. 

\section{An approximate formula for the precession angular speed.} \label{s4}

It is not hard to solve the equations of motion and find $\omega_p$ numerically. However, it is more intuitive to derive an analytical expression for it. We know that the function $\theta(\tau)$ in our case has the form
\begin{equation}
\theta(\tau)=\pi/2-\zeta P(\tau),
\end{equation}
\noindent where $P(\tau)$ is a periodic function that has a range of $[-1,1]$, and satisfies the initial condition $P(0)=0$. If we plug this solution in Eq.~\ref{phidot}, expand the resulting expression in powers of $\tau$ and use the initial condition of $P(\tau)$ we get
\begin{eqnarray}
\dot{\phi}(\tau)=\frac{\cl(r-2M)+2a\ce M}{r\Delta} +\mathcal{O}(\tau^2).
\end{eqnarray}
Integrating and using $\phi(0)=0$ give
\begin{eqnarray}
\phi(\tau)=\frac{\cl(r-2M)+2a\ce M}{r\Delta}\tau+\mathcal{O}(\tau^3).\label{phi3}
\end{eqnarray}

\nin Comparing with Eq.~\ref{phi2} we conclude that $\omega_p$ is contained in the  coefficient of the term linear in $\tau$ in Eq.~\ref{phi3} along with a contribution due to orbiting. This contribution, to a very good approximation, is given by the coefficient of the linear term with $a=0$. Because
\begin{equation}
\frac{\cl(r-2M)+2a\ce M}{r\Delta}\bigg\rvert_{a=0} = \frac{\cl}{r^2},
\end{equation}
\nin we can write an approximate expression for $\omega_p$ as
\begin{eqnarray}
\omega_p \approx \frac{\cl(r-2M)+2a\ce M}{r\Delta}-\frac{\cl}{r^2}.
\end{eqnarray}\label{formula}

\nin To see how good this formula is, we used it to reproduce the curves in Figs.~\ref{fig:omegap}~and~\ref{fig:omegan}. The results are represented by the dashed curves in the two figures. We can see that the dashed curves match the solid curves very well and the error in the formula for $\omega_p$ is not tangible in almost all cases. The error is noticeable, albeit still small, only when $a\approx M$ and $r$ is near $r_{_\text{{TISCO}}}$.

\section{Summary}\label{sum}

Circular orbits are the building blocks of accretion disks. We have shown that stable tilted circular orbits can exist around Kerr black hole. We studied the TISCOs and found that the tilt angle of the orbit increases the radius of the TISCO for prograde orbits and decreases it for retrograde orbits. When the tilt angle exceeds about $71^\circ$, the TISCO radius may correspond to two different values of the black hole's spin. The dependence of the last circular orbit on the tilt angle is qualitatively similar.

We studied the precession of stable circular orbits and found that they precess at a constant rate over the whole orbit in the direction of the black hole's spin. The precession rate was calculated numerically for a few representative cases. It was found to be almost independent of the tilt angle. The precession period was then discussed and calculated for these cases. We then derived a precise, approximate expression for the precession rate.   

The precession period for stellar mass black holes is typically few seconds. For supermassive black holes, the precession period is typically few days to few years. These findings show that tilted accretion disk models are viable for explaining quasi-period oscillations in astrophysical black hole systems.   

It would be interesting to study tilted circular orbits stability and precession around Kerr black holes in the presence of accretion disk electromagnetic fields. Such studied can lead to more astrophysically interesting findings.

\end{document}